\documentclass[aps,prl,twocolumn,superscriptaddress,floats,floatfix]{revtex4}
\usepackage{amssymb,amsmath}
\usepackage{graphics}
\usepackage{epstopdf}
\usepackage{color}
\usepackage{subfigure}
\usepackage{epsfig}
\usepackage{rotating}
\usepackage{float}
\usepackage{anyfontsize}
\usepackage{hyperref}

\begin{document}
\title{Rotating self-gravitating Bose-Einstein condensates with a crust: a minimal model for pulsar glitches}
\author{Akhilesh Kumar Verma}
\email{akhilesh@iisc.ac.in}
\affiliation{Centre for Condensed Matter Theory, Department of Physics, Indian Institute of Science, Bangalore 560012, India.}
\author{Rahul Pandit}
\email{rahul@iisc.ac.in}
\altaffiliation[\\ also at~]{Jawaharlal Nehru Centre For Advanced
Scientific Research, Jakkur, Bangalore, India.}
\affiliation{Centre for Condensed Matter Theory, Department of Physics, 
Indian Institute of Science, Bangalore 560012, India.} 
\author{Marc E. Brachet} 
\email{brachet@phys.ens.fr}
\affiliation{Laboratoire de Physique de 
l'\'{E}cole Normale Sup{\'e}rieure, 
ENS, Universit\'{e} PSL, CNRS, Sorbonne Universit\'{e}
Universit\'{e} de Paris,
24 Rue Lhomond, 75005 Paris, France}
\date{\today}
\begin{abstract}

We develop a minimal model for \textit{pulsar glitches} by introducing a
solid-crust potential in the three-dimensional (3D)
Gross-Pitaevskii-Poisson equation (GPPE), which we have used earlier to
study gravitationally bound Bose-Einstein Condensates (BECs), i.e.,
bosonic stars. In the absence of the crust potential, we show that, if
we rotate such a bosonic star, it is threaded by vortices.  We then
show, via extensive direct numerical simulations (DNSs), that the
interaction of these vortices with the crust potential yields (a)
stick-slip dynamics and (b) dynamical glitches.  We demonstrate that,
if enough momentum is transferred to the crust from the bosonic star,
then the vortices are expelled from the star, and the crust's angular
momentum $J_c$ exhibits features that can be interpreted naturally as
glitches.  From the time series of $J_c$, we compute the cumulative
probability distribution functions (CPDFs) of event sizes, event
durations, and waiting times. We show that these CPDFs have signatures
of self-organized criticality (SOC), which have been seen in
observations on pulsar glitches. 	

\end{abstract}
\keywords{Superfluidity; quantum fluids; quantum vortices; pulsar glitches}
\maketitle

Rotating magnetized neutron stars~\cite{Melatos2015,Basu_2018}, or
\textit{pulsars}, display \textit{glitches}, which are sudden increases of
their rotational frequencies. These observations have a long
history~\cite{radhakrishnan1969detection,boynton1969precision,manchester_2017};
and they indicate that glitches are associated with the transfer of angular
momentum, which is carried by quantum vortices in the superfluid interior, to
the solid crust, in the outer layers of the pulsar. This transfer occurs
because of vortex-crust interactions, as suggested in
Refs.~\cite{Ruderman69,baym1969spin}. The quantitative modeling of pulsar
glitches is complex, so different models have been
suggested~\cite{Melatos2015,Sidery2010}: some involve avalanches of superfluid
vortices~\cite{khomenko_haskell_2018}; other mainstream models are based on
neutron superfluidity in the crust~\cite{PhysRevLett.119.062701}.

Neutron Cooper pairs~\cite{Migdal59}, which comprise a major component of the
nuclear matter in a pulsar, form a superfluid. Therefore,
Refs.~\cite{Warsz_2011,PhysRevB.85.104503} have proposed simply to model
this superfluid by using the two-dimensional (2D) Gross-Pitaevskii equation
(GPE); in addition, they have included an externally imposed potential or
\textit{container} and a \textit{pinning} potential for the crust.
However, pulsars are three dimensional (3D); and gravitational effects are
important on stellar scales. It is important, therefore, to account for these
crucial features in a model for pulsars and the glitches they exhibit. In this
Letter, we construct a \textit{natural, minimal model} for pulsar glitches by
(a) accounting for gravitational effects via the Gross-Pitaevskii-Poisson equation
(GPPE) for a self-gravitating superfluid (see, e.g.,
Refs.~\cite{Chavanis1,Chavanis2,Akhilesh2019} for a non-rotating bosonic star)
and (b) including \textit{rotation} and an \textit{interacting} solid crust. 

We carry out extensive pseudospectral direct numerical simulations (DNSs) to
show that our minimal model yields pulsar glitches with properties that are
akin to those seen in observations. In particular, the time series of the
angular momentum  $J_c$ of the crust shows the hallmarks of self-organized
criticality
(SOC)~\cite{bak1987self,carlson1989properties,jensen1998self,turcotte1999self,morley1993scaling,melatos2008avalanche,aschwanden2013self,aschwanden201625}.
To obtain these results, we develop a sophisticated algorithm to find the
ground state of the GPPE, with rotation: this alogorithm uses an ancilliary
advective real Ginzburg-Landau-Poisson equation (ARGLPE), an imaginary-time
version of the GPPE. The resulting ground states \textit{contain vortices} and
yield uniformly rotating solutions of the GPPE. When we include the crust and its
dynamics, we find a transfer of the angular momentum from the star to the
crust, where it is dissipated by friction; if this transfer is large enough,
vortices move outwards and glitches are observed.

Self-gravitating GPPE superfluids are described by a complex wave function
$\psi({\bf x},t)$, governed by the following partial differential equation (PDE): 
\begin{eqnarray} 	
i \hbar \partial_t \psi  &=& -\frac{\hbar^2}{2m} \nabla^2 \psi  +
\left[V_\theta+G \Phi +g |\psi|^2\right] \psi ; \label{eq:GPPE}\\
\nabla^2 \Phi &=& |\psi|^2  - < |\psi|^2> ; \label{eq:Poisson}
\end{eqnarray} 	
here, $m$ is the mass of the bosons, $n=|\psi|^2$ their number density, $G =
4\pi G_{N}m^2$ ($G_N$ denotes Newton's gravitational constant), and $g=4 \pi a
\hbar^2/m$, with $a$ the $s$-wave scattering length~\footnote{Note that the
subtraction of the mean density on the right-hand side (RHS) of the Poisson
equation~\ref{eq:Poisson} can be justified either by taking into account the
cosmological expansion~\cite{Peebles,falco2013} or by defining a Newtonian
cosmological constant~\cite{KIESSLING2003}.} We describe the dynamics of the pulsar's
solid crust by a single polar angle $\theta$, which evolves as follows:
\begin{eqnarray} 	
I_c \frac{d^2\theta}{dt^2}&=&\frac{1}{N}\int d^3x \partial_\theta V_\theta |\psi |^2 -\alpha \frac{d\theta}{dt};
\label{eq:theta} \\
	V_\theta({\bf r}_p) &=& V_0 \exp(-\frac{|{\bf r}_p |-r_{\rm crust}}{\Delta r_{\rm 
crust}})^2 \tilde V(x_\theta,y_\theta) ;
\label{eq:Vtheta}
\end{eqnarray} 
$I_c$ and $V_\theta$ denote, respectively, the moment of inertia of the crust
and the crust potential; $\alpha$ controls the frictional slowing down of the
rotation of the crust, with $\sqrt{I_c/\alpha}$ the crust-friction decay time;
for specificity, we choose $\tilde V(x_\theta,y_\theta) = 3+\cos(n_{\rm
crust}x_\theta)+\cos(n_{\rm crust}y_\theta))$, with $x_\theta = \cos(\theta)x_p
+ \sin(\theta)y_p$ and $y_\theta = -\sin(\theta)x_p + \cos(\theta)y_p$; here,
$n_{\rm crust}$ determines the number of pinning sites in the crust potential,
$r_{\rm crust}$ is the radius at which this potential is a maximum, and $\Delta
r_{\rm crust}$ determine the thickness of the crust.  We use periodic boundary
conditions (PBCs), so, to obtain a periodic version of the angular momentum $J_z$ of
the GPPE condensate, we use the $\pi$-centered, $2 \pi$-periodic coordinates:
${\bf r}_p=(x_p,y_p,z_p)$, with $x_p=-\sum_1^{10} \exp{(-\frac {16 }{100}n^2)}
(-1)^n \frac {\sin(n (x-\pi)}{n}$, $y_p=-\sum_1^{10} \exp{(-\frac {16
}{100}n^2)} (-1)^n \frac {\sin(n (y-\pi)}{n}$, and $z_p = \pi$  ($x_p$ and
$y_p$ are nearly linear near the center because they are truncated Fourier
expansions of saw-tooth waves). 

The crust \textit{acts} on the GPPE superfluid \textit{and also reacts} to
it~\footnote{The first term on the RHS of Eq.~\eqref{eq:theta} can be obtained
from the GPPE Lagrangian augmented with the crust rotational energy
$\frac{I_c}{2} (\frac{d\theta}{dt})^2$; the variation of this Lagrangian with
respect to $\bar \psi$ yields Eq.~\eqref{eq:GPPE}; and the variation with
respect to $\theta$ yields the first term on the RHS. of Eq.~\eqref{eq:theta}
(see Ref.~\cite{Vishwanath2016} for a similar procedure involving active
particles in the Gross-Pitaevskii equation with Newtonian particles).} .  If
$\alpha = 0$, the GPPE-crust coupled system \eqref{eq:theta} and \eqref{Eq:TGPPEphys}
(or the Lagrangian from which it is derived) obeys the
following global conservation laws: (a) Rotational invariance, \textit{in the
whole space} $\mathbb{R}^3$, which leads to the conservation of the total
angular momentum $J=J_c+J_z$, with $J_c=I_c \frac{d\theta}{dt}$ the crust
angular momentum and $J_z=\int d^3 x  \bar \psi ({\bf \hat e}_z \times {\bf r})
\cdot(- i \hbar \nabla) \psi$. (b) Time-translation invariance,  by virtue of
which the total energy $E_{tot}=E+E_c$ is also conserved; here,
$E_c=\frac{1}{2} (\frac{d\theta}{dt})^2$ is the rotational energy of the crust
and $E$ is the energy of the GPPE system, which we rewrite as 
\begin{equation}
i\hbar\frac{\partial\psi}{\partial t} = - \frac{\hbar^2}{2m} {\bf \nabla}^2 \psi +  [V_\theta+(G {\bf \nabla}^{-2}+g)|\psi|^2]\psi .
\label{Eq:TGPPEphys}
\end{equation}
Equation~(\ref{Eq:TGPPEphys}) conserves the number of particles $N=\int d^3 x
|\psi|^2$ and (for time-independent $\theta$) the GPPE energy
$E=E_{kq}+E_{int}+E_{G}+E_{V}$, where $E_{kq}=\frac{\hbar^2}{2m} \int d^3 x
|{\bf \nabla} \psi |^2$, $E_{int}=\frac{g}{2} \int d^3 x (|\psi|^2)^2 $,
$E_{G}=\frac{G}{2} \int d^3 x (|\psi|^2) {\bf \nabla}^{-2} (|\psi|^2)$ and
$E_{V}=\int d^3 x |\psi|^2 V_\theta$.  For $V_\theta=0$, the momentum
${\bf P}=\frac{i\hbar}{2}\int d^3x\left( \psi {\bf \nabla}\overline{\psi} -
\overline{\psi} {\bf \nabla}\psi\right)$ is also conserved. 

We solve the GPPE~\ref{Eq:TGPPEphys} by using a 3D Fourier pseudospectral
method~\cite{Krstulovic11,vmrnjp13,Got-Ors}, with $\psi(x)=\sum_{|{\bf
k}|<k_{\rm max}} \hat \psi_{\bf k} \exp(i {\bf k}\cdot{\bf x})$ and $k_{\rm
max}=[{\cal N}/3]$, where $\cal{N}$ is the resolution and $[\cdot]$ denotes the
integer part.  In the absence of the crust potential, friction, and of
rotation, we obtain the conventional GPPE [see Ref.~\cite{Akhilesh2019}, where
we discuss the gravitational Jeans instability and the pseudospectral method
(see also the Supplemental Material~\cite{supmat})]. In all $3$ spatial
directions, our DNS uses $2 \pi$-periodic PBCs, which we also use for $J_z$ and
$V_\theta$; so, even if $\alpha=0$, the conservation of the total angular
momentum holds only approximately, for the system does not have rotational
invariance in $\mathbb{R}^3$. 


We first obtain uniformly rotating states for the GPPE with rotational speed 
$\Omega$ by solving the imaginary-time equation
\begin{equation}
\hbar \partial_t \psi =-\frac {\delta}{\delta \bar \psi} (E-\Omega J_z-\mu
N-\lambda (\frac{\bf P}{N_0 m})^2).
\end{equation}
Rotational ground states are minima of $E-\Omega J_z$; $\mu$, the chemical potential,
and $\lambda$ are Lagrange multipliers; at each time step, we tune $\mu$ to keep
the boson number fixed; and we choose a large value of $\lambda$ so that 
${\bf P}$ is small. We now obtain the following advective real Ginzburg-Landau 
equation (ARGLPE):
\begin{eqnarray}
\nonumber \hbar\frac{\partial\psi}{\partial t}&=& \frac{\hbar^2}{2m} {\bf \nabla}^2 \psi + \mu \psi  -  [V_\theta+(G {\bf \nabla}^{-2}+g)|\psi|^2]\psi  \\
&&- i \hbar (\Omega {\bf \hat e}_z \times {\bf r}_p-\lambda \frac {\bf P}{N_0 m}) \cdot \nabla \psi \label{eq:ARGL},\hspace{3mm}
\end{eqnarray}
which we solve to obtain the rotational ground states (minima) mentioned above;
to stabilise this minimization procedure, we reset the center of mass ${\bf
r}_{cm}=\int d^3x {\bf r}_p |\psi|^2/N$ to $(\pi,\pi,0)$, after each time step.

\begin{figure}
\centering	
\resizebox{\linewidth}{!}{
\includegraphics[width=5.8cm]{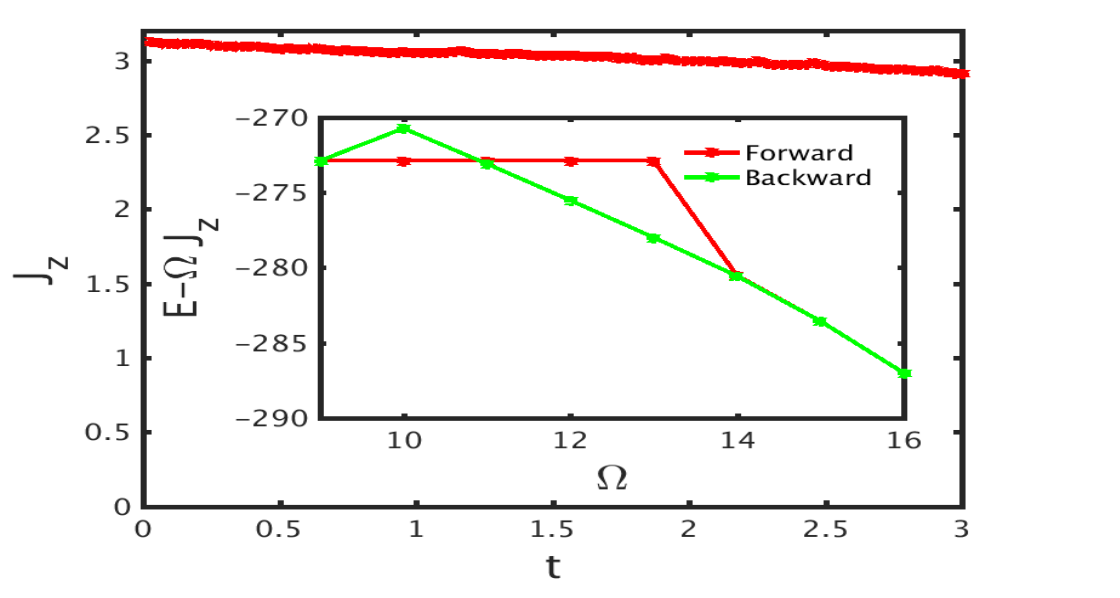}}
\caption{Illustrative plot of the GPPE condensate angular momentum $J_z$ vs
time $t$; $J_z$ is not conserved exactly (decay of
$\simeq 1\%$ per turn) because rotational symmetry holds only
approximately if we use PBCs. Parameters: ${\mathcal N}=64$, $g = 5$, $G = 50$, 
and $\Omega = 14$; the system has $4$ vortices. Inset: Plot of the ARGLPE-converged value 
(see text) of $E-\Omega J_z$ vs the rotation speed $\Omega$; the differences 
between scans with increasing (red) and decreasing (green) $\Omega$ indicate
hysteresis; the system has no vortices along the horizontal part of the
red line and $4$ vortices along the green one. }
\label{fig:ang_moment}
\end{figure}

\begin{figure}
\centering
\resizebox{\linewidth}{!}{
\includegraphics[scale=0.1]{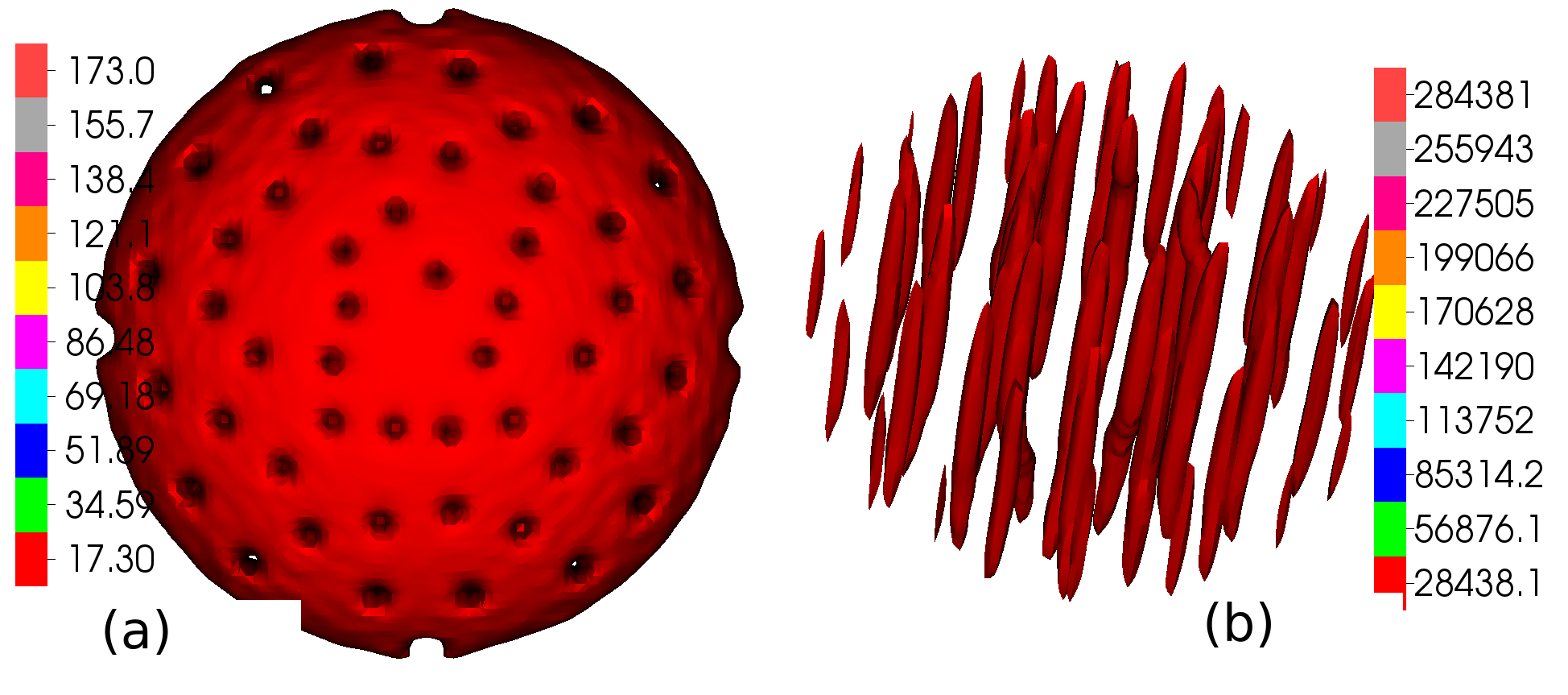}}
\caption{Isosurface plots illustrating a rotating, compact object with vortices 
obtained via the ARGLPE (see text): (a) Isosurfaces of the boson density (top view); 
for the spatiotemporal evolution of these isosurfaces see the video S1 in the 
Supplemental Material~\cite{supmat}; and 
(b) isosurfaces of $(\nabla \times (\rho v))^2$ (side view); here,
${\mathcal N}=256$, $G=800$, $g=80 $, and $\Omega = 60$.}\label{fig:rotating_star}
\end{figure}

We obtain rotational ($\Omega \neq 0$) states by integrating
Eq.~\eqref{eq:ARGL} until we get convergence; given our initial data, the
system contains $N$ bosons.  For $\Omega = 0$ and $V_\theta =0$, the solution
of the ARGLPE~\eqref{eq:ARGL} converges, at large times, to ground states
$\psi_0$ that are spherically symmetric, compact objects of radius $R$. This
radius can be estimated by using a variational ansatz~\cite{Eby2018} or it can
be computed numerically at both zero temperature ($T =
0$)~\cite{Chavanis1,Chavanis2} and finite temperature ($T >
0$)~\cite{Akhilesh2019}. We start from the $T=0$ state of
Ref.~\cite{Akhilesh2019}; and then we increase the value of $\Omega$ in steps
of $1$; at each such increase in the value of $\Omega$, we use the converged
ARGLPE state, from the previous value of $\Omega$, as the initial data. In the
inset of Fig.~\ref{fig:ang_moment} we plot the  ARGLPE-converged values of
$E-\Omega J_z$ versus $\Omega$; from this plot it is apparent that the
non-rotating state, with no vortices, loses its stability around $\Omega \simeq
14$, to a state with $4$ vortices. We show scans in which $\Omega$ increases
(red lines) and decreases (green lines). The differences between these scans
indicate that this system exhibits hysteresis: It has no vortices along the
horizontal part of the red line and $4$ vortices along the green ones; and it
goes from the $4$-vortex branch back to the $0$-vortex branch at $\Omega \simeq
9$.  In Fig.~\ref{fig:ang_moment} we give an illustrative plot of
$J_z$ versus time $t$ for the GPPE evolution of the $4$-vortex state at
$\Omega=14$; we see that $J_z$ is not conserved
perfectly, but it decreases slightly ($\simeq 1\%$ per turn). $J_z$ is
not conserved exactly because rotational symmetry holds only approximately,
given our PBCs; the closer we are to the center of our
computational domain the better is this rotational symmetry.  We show, in Figs.
\ref{fig:rotating_star}(a) and (b), respectively, isosurface plots of the boson
density (top view) and of $(\nabla \times (\rho v))^2$ (side view) for the
converged solution of ARGLPE, with ${\mathcal N}=256$, $G=800$, $g=80 $, and
$\Omega = 60$. Clearly, our ARGLPE-based algorithm yields
rotating, self-gravitating Bose-Einstein condensates (BECs), with vortices in
the GPPE; this has not been possible hitherto~\footnote{For example, such
states were not obtained in Ref.~\cite{PhysRevD.91.044041}, probably because
pure GPPE evolution was used, without the benefit of ARGLPE-based
minimization.}.

\begin{figure*}[ht]
\centering	
\resizebox{\linewidth}{!}{
\includegraphics[scale=0.1]{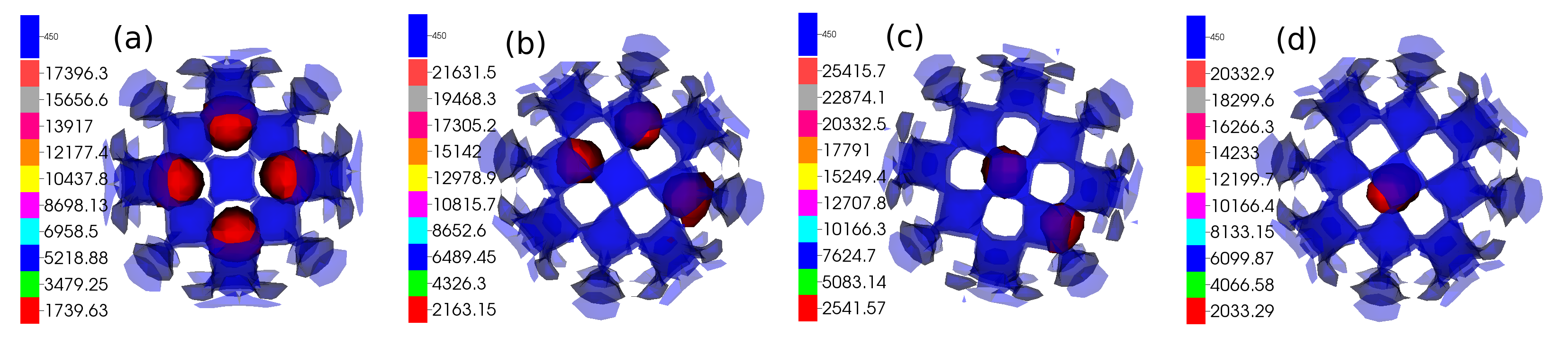}}

\caption{Plots of crust-potential isosurfaces in blue, with $V_{\theta}=450$, and
of ten-level isosurfaces of $(\nabla \times (\rho v))^2$, from our DNS of the 
GPPE, for the representative parameter values  $V_0 = 180$, $n_{\rm crust}=12$, 
$I_c = 0.01$, $r_{\rm crust} = 1.0$, $\Delta r_{\rm crust} = 0.15$, $\Omega_0 = 14$, and 
$\alpha = 0.007$ (as in Fig.~\ref{fig:moment_crust} below) at times (a) $t=0.06$, 
(b) $t=6.48$, (c) $t=7.38$, and (d) $t=9.72$. For the spatiotemporal evolution of 
these isosurfaces see the video S2 in the Supplemental Material~\cite{supmat}.}

\label{fig:glitch_plot}
\end{figure*}

\begin{figure*}[ht]
\centering
\resizebox{\linewidth}{!}{
\includegraphics[scale=1.0]{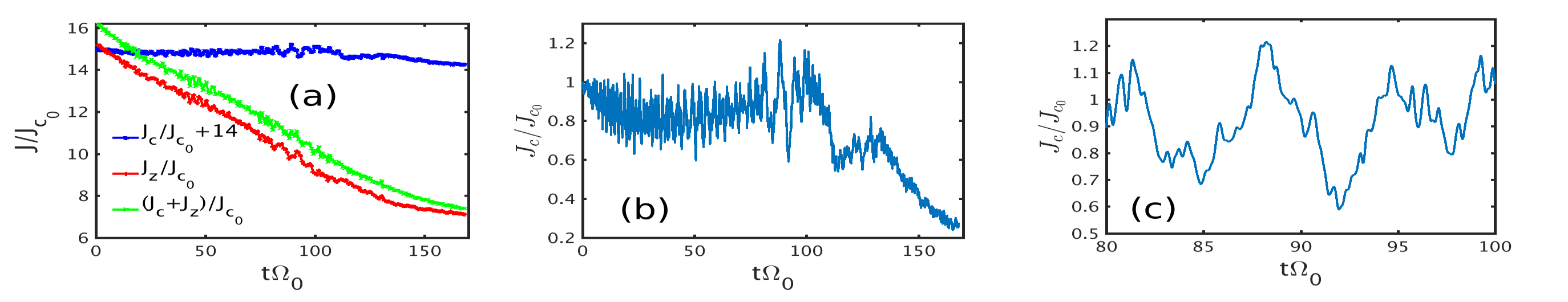}}

\caption{(a) Plots, versus the scaled time $t\Omega_0$, of $(J_c/J_{c_0} +14)$
[blue curve], $J_z/J_{c_0}$ [red curve], and $(J_c + J_z)/J_{c_0}$ [green
curve], where $J_{c_0}$ is the initial angular momentum of the crust, for
the representative parameter values given in Fig.~\ref{fig:glitch_plot}
above. Expanded plots of $J_c/J_{c_0}$ for (b) $0 < t\Omega_0 \lesssim
170$ and (c) $80 \leq t\Omega_0 \leq 100$ .}
\label{fig:moment_crust}
\end{figure*}

We turn now to the GPPE~\eqref{Eq:TGPPEphys} coupled with the crust-rotation
equation~\eqref{eq:theta}. In Figs.~\ref{fig:glitch_plot} (a)-(d), we plot the
crust-potential isosurface in blue, with $V_{\theta} = 450$, and ten-level
isosurfaces of $(\nabla \times (\rho v))^2$ for a representative set of
parameters and the times (a) $t=0.06$, (b) $t=6.48$, (c) $t=7.38$, and (d)
$t=9.72$. \textit{On average}, the crust gains angular momentum from the
superfluid, hence, at long times, these vortices move outwards after losing
enough angular momentum to the crust (compare
Figs.~\ref{fig:glitch_plot} (a) and (d)). However, the time series of the
condensate and crust angular momenta, $J_z$ and $J_c$, respectively, are
complicated, and, as we show below, they display the signatures of
SOC~\cite{bak1987self,carlson1989properties,jensen1998self,turcotte1999self,morley1993scaling,melatos2008avalanche,aschwanden2013self,aschwanden201625}.  We
illustrate this temporal evolution in Fig.\ref{fig:moment_crust},
for the representative parameter values  $V_0 = 180$, $n_{\rm crust}=12$, $I_c
= 0.01$, $r_{\rm crust} = 1.0$, $\Delta r_{\rm crust} = 0.15$, $\Omega_0 = 14$, and
$\alpha = 0.007$.  In Fig.~\ref{fig:moment_crust}(a) we plot, versus the scaled
time $t\Omega_0$, $(J_c/J_{c_0} - 14)$ [blue curve], $J_z/J_{c_0}$ [red curve], and
$(J_c + J_z)/J_{c_0}$ [green curve], where $J_{c_0}$ is the initial angular
momentum of the crust. This figure shows that, if we neglect the overall,
gentle decay of the total angular momentum~\footnote{We have noted above that, because we
use $2 \pi$-periodic coordinates to define $J_z$, the conservation of angular
momentum is only approximate, \textit{even if} $\alpha=0$; also, in neutron
stars, the crust angular momentum is only a few percent of the total angular
momentum~\cite{Basu_2018}.}, fluctuations of $J_c$ compensate for those in
$J_z$. In Figs.~\ref{fig:moment_crust}(b) and (c) we show expanded plots of
$J_c/J_{c_0}$ for $0 < t\Omega_0 \lesssim 170$ and $80 \leq t\Omega_0 \leq 100$,
respectively, to illustrate the irregular nature of the time series of the
angular momentum of the crust.

\begin{figure*}[ht]
\centering	
\resizebox{\linewidth}{!}{
\includegraphics[scale=0.5]{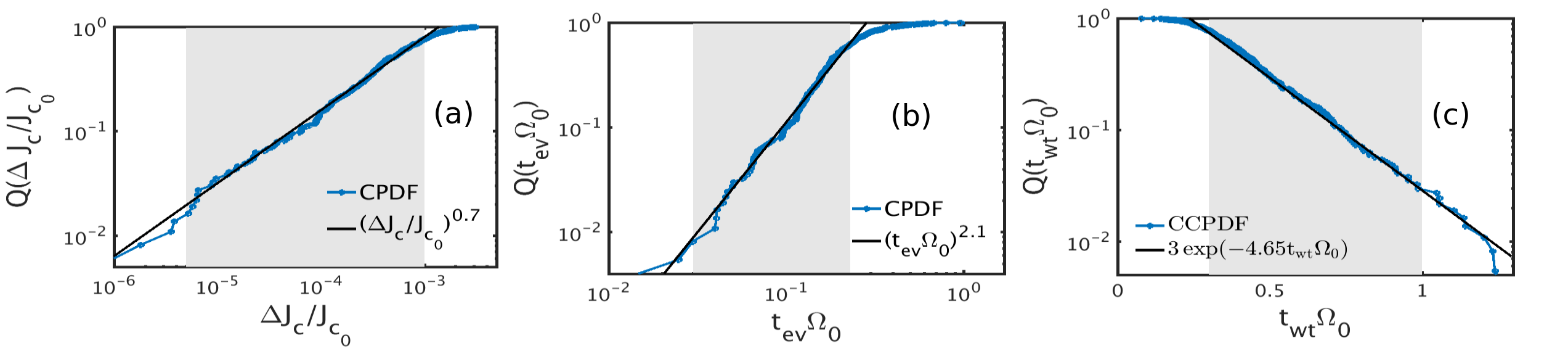}}
\caption{Log-log (base 10) plots of: (a) the CPDF $Q(\Delta J_c/J_{c_0})$ of 
the event size; and (b) the CPDF $Q(t_{\rm ed}\Omega_0)$ of the event duration. 
(c) Semi-log (base 10) plot of the CPDF $Q(t_{\rm wt}\Omega_0)$. $J_{c_0}$ and
$\Omega_0$ are, respectively, the initial angular momentum and angular velocity 
of the crust. Our DNS data are shown in blue; the black lines show fits 
(power-law or exponential) to these data in the shaded gray regions in the 
plots.}
\label{fig:soc_plot}
\end{figure*}

From the time series $J_c(t)$ (Figs.~\ref{fig:moment_crust}(a)-(c)), we see
that the crust can lose (gain) angular momentum to (from) the superfluid,
because  vortices stick to (slip from) the crust. To characterize the
statistical properties of these stick-slip events, we calculate the gain
$\Delta J_c$ in the crust angular momentum, between successive minima and
maxima of $J_c(t)$; we call $\Delta J_c$ the event size; we scale it by
$J_{c_0}$.  In Fig.~\ref{fig:soc_plot}(a) we present a log-log (base 10) plot of
the cumulative probability distribution function (CPDF) $Q(\Delta
J_c/J_{c_0})$; this yields the power-law behavior $Q(\Delta J_c/J_{c_0}) \sim
(\Delta J_c/J_{c_0})^{\beta}$, for the part of the CPDF that lies in the region
shaded gray; thus, the probability distribution function (PDF) $P(\Delta
J_c/J_{c_0})\sim(\Delta J_c/J_{c_0})^{\beta-1}$; by fitting the CPDF in the
gray region, we find $\beta \simeq 0.7$. 

Next, we calculate the event-duration time $t_{\rm ed}$, i.e., the time
difference between successive minima and maxima of $J_c(t)$, and thence the
CPDF  $Q(t_{\rm ed}\Omega_0)$ (log-log, base 10, plot in
Fig.~\ref{fig:soc_plot}(b)). This plot shows, in the shaded gray region, that
$Q(t_{\rm ed}\Omega_0) \sim (t_{\rm ed}\Omega_0)^{\gamma}$, with $\gamma \simeq
2.1$. Clearly, the PDF $P(t_{\rm ed}\Omega_0) \sim (t_{\rm ed}
\Omega_0)^{\gamma-1}$, in this region. 

Finally, we compute the CPDF $Q(t_{\rm wt}\Omega_0)$ of the waiting time
$t_{\rm wt}$, i.e., the time difference between two successive maxima. From
the shaded gray region in the semi-log (base 10) plot in Fig.\ref{fig:soc_plot}(c),
we observe the exponential form $Q(t_{\rm ed}\Omega_0) \sim \exp(-t_{\rm ed}\Omega_0)$. 

These power-law behaviors of $Q(\Delta J_c)$ and $Q(t_{\rm ed})$ and
the exponential tail of $Q(t_{\rm wt})$ together show that the stick-slip motion,
between superfluid vortices and the crust, yields the time series $J_c(t)$, which has all
the signatures of SOC found in measurements of pulsar
glitches~\cite{morley1993scaling,melatos2008avalanche,aschwanden2013self,aschwanden201625}.
Some classes of pulsars exhibit a glitch-size PDF of the type we have obtained
in our model; in particular, they show a power-law behavior in this PDF, over a
certain range of sizes; and the power-law exponents lie in the range $-0.13
\lesssim -(\beta-1) \lesssim 2.4$. For the pulsar PSR J
1825-0935~\cite{melatos2008avalanche}, the exponent for the power-law
glitch-size PDF is $\simeq 0.36$, nearly the same as our calculated
exponent. In many pulsars, including PSR J 1825-0935, the
waiting-time PDF has an exponential tail; this is in agreement with our
result~\cite{melatos2008avalanche}.

We have shown that uniformly rotating vortex-containing gravitationally-bound
solutions of the GPPE can be generated by starting the evolution from initial
data obtained by integrating to convergence the (imaginary-time)
ARGLPE~\eqref{eq:ARGL}. We have built on this GPPE and introduced
a minimal model, with a single, angular, dynamical variable for a solid crust
coupled with a rotating GPPE star. We have demonstrated that this model
exhibits stick-slip dynamics, whose statistical properties we have
characterized by computing the event-size and event-duration CPDFs
$Q(\Delta J_c/J_{c_0})$ and $Q(t_{\rm ed}\Omega_0)$, which show power-law
forms, and the waiting-time  CPDF $Q(t_{\rm wt}\Omega_0)$, which exhibits an
exponential tail. These SOC-type desiderata are in consonance with
measurements on a class of pulsars~\cite{melatos2008avalanche}.

We plan to study pulsar-glitch models that are more realistic
than our minimal model. Examples include models with (a) a solid crust with $6$
degrees of freedom, $3$ rotational and  $3$ translational, instead of only one
angle of rotation, or (b) a superconducting component with magnetic flux tubes.
We expect that such generalizations of our minimal model should help us to
undertand all the types of statistical properties that are displayed by pulsar
glitches in different
pulsars~\cite{morley1993scaling,melatos2008avalanche,aschwanden2013self,aschwanden201625,espinoza2011study,carlin2019generating,eya2019distributions}.

\begin{acknowledgments}
We thank DST (IN), CSIR (IN), and the Indo-French Center for 
Applied Mathematics (IFCAM) for their support and SERC (IISc) for computational
resources.
\end{acknowledgments}


\begin{center}
\pagebreak
\widetext
\textbf{\large Supplemental Materials:Rotating self-gravitating Bose-Einstein condensates with a crust: a minimal model for
pulsar glitches}
\author{Akhilesh Kumar Verma}
\email{akhilesh@iisc.ac.in}
\affiliation{Centre for Condensed Matter Theory, Department of Physics, Indian
Institute of Science, Bangalore 560012, India}
\author{Rahul Pandit}
\email{rahul@iisc.ac.in}
\altaffiliation[\\ also at~]{Jawaharlal Nehru Centre For Advanced
Scientific Research, Jakkur, Bangalore, India}
\affiliation{Centre for Condensed Matter Theory, Department of Physics,
Indian Institute of Science, Bangalore 560012, India}
\author{Marc E. Brachet}
\email{brachet@physique.ens.fr}
\affiliation{Laboratoire de Physique de
l'\'{E}cole Normale Sup{\'e}rieure,
ENS, Universit\'{e} PSL, CNRS, Sorbonne Universit\'{e}
Universit\'{e} de Paris, F-75005 Paris, France}
\end{center}
\setcounter{equation}{0}
\setcounter{figure}{0}
\setcounter{table}{0}
\setcounter{page}{1}
\makeatletter
\renewcommand{\theequation}{S\arabic{equation}}
\renewcommand{\thefigure}{S\arabic{figure}}
\renewcommand{\bibnumfmt}[1]{[S#1]}
\renewcommand{\citenumfont}[1]{S#1}

In this Supplemental Matrial we provide details of our direct numerical
simulation and some videos from our DNSs of the advective real Ginzburg-Landau-
Poisson equation (ARGLPE) and the Gross-Pitaevskii-Poisson equation (GPPE) with
a crust potential.

{\bf Our DNS}

We solve the GPPE Eq.(1) and (2) of the main text by using a 3D Fourier pseudospectral
method~\cite{Krstulovic11,vmrnjp13,Got-Ors}, with $\psi(x)=\sum_{|{\bf
k}|<k_{\rm max}} \hat \psi_{\bf k} \exp(i {\bf k}\cdot{\bf x})$ and $k_{\rm
max}=[{\cal N}/3]$, where $\cal{N}$ is the resolution and $[\cdot]$ denotes the
integer part.  In the absence of the crust potential, friction, and of
rotation, we obtain the conventional GPPE [see Ref.~\cite{Akhilesh2019}, where
we describe the large-scale gravitational Jeans instability.
In all $3$ spatial directions, our DNS uses $2
\pi$-periodic boundary conditions (PBCs), which we also use to define $J_z$ and
$V_\theta$; so, even if $\alpha=0$, the conservation of the total angular
momentum holds only approximately, for the system does not have rotational
invariance in $\mathbb{R}^3$.

We use the units that we have employed in Ref.~\cite{Akhilesh2019}, where we
have shown that non-rotating, compact, gravitationally bound objects have a
radius of gyration, $R_G = \sqrt{\frac{\int_V \rho(r) r^2 d{\bf r}}{\int_V
\rho(r)d{\bf r}}}$ that is of the order of the length scale $R_a=\sqrt{\frac{a
\hbar^2}{G_N m^3}}=\sqrt{\frac{g}{G}}$. Our DNS employs a $(2 \pi)^3$ periodic
computational box, so we normalize $\psi$ such that $N=N_0=(2\pi)^3$; and, in
DNS time and length units, $\hbar = 1$ and $m = 1$. For time marching we use
the fourth-order Runge-Kutta scheme.

{\bf Videos from our DNSs}

\begin{itemize}
\item The spatiotemporal evolution of $|\psi({\bf x},t)|^2$ from our ARGLPE
study, for ${\mathcal N}=256$, $G=800$ $g=80 $, and $\Omega = 60$ (cf.,
Fig. $2$ in the main paper) is given in the video V1.
\item The spatiotemporal evolution of the crust-potential isosurfaces in blue,
with $V_{\theta}=450$, and isosurfaces of $(\nabla \times (\rho v))^2$,
from our DNS of the GPPE, for the representative parameter
values  $V_0 = 180$, $n_{\rm crust}=12$, $I_c = 0.01$, $r_{\rm
crust} = 1.0$, $\Delta r_{c} = 0.15$, $\Omega_0 = 14$, and
$\alpha = 0.007$ (cf., Fig. $3$ in the main paper) is given in
the video V2.
\end{itemize}

\end{document}